\documentclass[twocolumn,english,prl,floatfix,citeautoscript,nofootinbib,superscriptaddress]{revtex4}
\usepackage{amsbsy}
\usepackage{latexsym,epsfig,graphicx}
\usepackage{dcolumn}
\usepackage{subfigure}
\usepackage{comment}
\usepackage{color}
\usepackage[colorlinks,urlcolor=blue,citecolor=blue]{hyperref}
\usepackage{amstext}
\usepackage{amssymb}
\usepackage{setspace}

\begin{document}

\title{Topological photonic orbital angular momentum switch}
\author{Xi-Wang Luo}
\affiliation{Key Laboratory of Quantum Information, University of Science and Technology
of China, Hefei, Anhui 230026, China}
\affiliation{Department of Physics, The University of Texas at Dallas, Richardson, Texas
75080-3021, USA}
\author{Chuanwei Zhang}
\thanks{Email: chuanwei.zhang@utdallas.edu}
\affiliation{Department of Physics, The University of Texas at Dallas, Richardson, Texas
75080-3021, USA}
\author{Guang-Can Guo}
\affiliation{Key Laboratory of Quantum Information, University of Science and Technology
of China, Hefei, Anhui 230026, China}
\affiliation{Synergetic Innovation Center of Quantum Information and Quantum Physics,
University of Science and Technology of China, Hefei, Anhui 230026, China}
\author{Zheng-Wei Zhou}
\thanks{Email: zwzhou@ustc.edu.cn}
\affiliation{Key Laboratory of Quantum Information, University of Science and Technology
of China, Hefei, Anhui 230026, China}
\affiliation{Synergetic Innovation Center of Quantum Information and Quantum Physics,
University of Science and Technology of China, Hefei, Anhui 230026, China}

\begin{abstract}
The large number of available orbital angular momentum (OAM) states of
photons provides a unique resource for many important applications in
quantum information and optical communications. However, conventional OAM
switching devices usually rely on precise parameter control and are limited
by slow switching rate and low efficiency. Here we propose a robust, fast
and efficient photonic OAM switch device based on a topological process,
where photons are adiabatically pumped to a target OAM state on demand. Such
topological OAM pumping can be realized through manipulating photons in a
few degenerate main cavities and involves only a limited number of optical
elements. A large change of OAM at $\sim 10^{q}$ can be realized with only $%
q $ degenerate main cavities and at most $5q$ pumping cycles. The
topological photonic OAM switch may become a powerful device for broad
applications in many different fields and motivate novel topological design of
conventional optical devices.
\end{abstract}

\maketitle

\section{Introduction}
Discrete degrees of freedom, such as charge, spin,
valleys, \textit{etc.}, play a crucial role in many information encoding and
device applications~\cite{nielsen2002quantum, gisin2002quantum,
culcer2012valley, andersen2015hybrid}. In this context, a fundamental degree
of freedom of photons, the orbital angular momentum (OAM), possesses a
unique property that an infinite number of distinctive OAM states are
available~\cite{allen1992orbital, molina2007twisted, yao2011orbital}. This
unique property makes photonic OAM very attractive for various applications
in optical communication~\cite{malik2016multi, barreiro2008beating,
wang2012terabit}, quantum simulation~\cite{cardano2015quantum,
cardano2016statistical, luo2015quantum, zhou2017dynamically}, quantum information~\cite%
{luo2017synthetic, fickler2012quantum, fickler2014interface, wang2015quantum}, and quantum
cryptography (e.g., key distribution)~\cite{groblacher2006experimental,
mafu2013higher, vallone2014free, mirhosseini2015high}. To fully utilize
these applications, a tunable device that can rapidly and robustly switch
between different OAM modes on demand is therefore highly desirable.
However, many conventional OAM switching devices rely on precise parameter
control and are usually limited by slow switching rates ($\sim $kHz)~\cite%
{thalhammer2013speeding, mirhosseini2013rapid} or low purity and efficiency~%
\cite{radwell2014high, strain2014fast}, and limited number of usable OAM
modes \cite{marrucci2006optical, slussarenko2011efficient}.

The photonic OAM is a discrete degree of freedom that characterizes
the topological charge (i.e., the winding of the azimuthal phase)
of a photon field with cylindrical symmetry. Therefore a natural
question is whether a topological process can be designed to create a robust
photonic OAM switching device with high performance. Recently, the study of
topological photonics has become one frontier direction in optical physics
with the major focus on modulating photon propagation through topological
edge states~\cite{haldane2008possible, hafezi2011robust, fang2012realizing,
lin2014light, lu2014topological, yuan2016photonic, lin2016photonic}, while
practical topological photonic devices for on-demand switching of photon
internal degrees of freedom are still largely lacking.

In this paper, we propose a practical photonic OAM switching device through
the topological adiabatic pumping of OAM states, which is robust (immune to
small perturbations in system parameters), fast ($\sim$ MHz switching rate),
efficient ($\sim 90\%$ efficiency and $\sim 100\%$ purity in principle), and
highly tunable (on demand switch for high-OAM states). Topological pumping
was initially proposed by Thouless~\cite{thouless1983quantization} for a
periodically time-varying system, and has been recently realized in
ultra-cold atom optical lattices \cite{lohse2015thouless,
nakajima2016topological} and coupled optical waveguides \cite%
{kraus2012topological, verbin2015topological}, where particles are
adiabatically pumped in real space lattices. Such \textquotedblleft Thouless
pumps\textquotedblright\ along particle's internal degrees of freedom have
not been well explored, although atomic hyperfine states have been utilized
as a synthetic dimension for studying quantum Hall effects \cite%
{celi2014synthetic, mancini2015observation, stuhl2015visualizing}.

Here we show how to realize Thouless pumps in the synthetic OAM space of
photons where large numbers of OAM modes are available,
which provides the basis for engineering a topological photonic OAM switch.
Essentially, we propose a scheme to realize a tunable double well OAM
lattice (two lattice sites in each unit cell) by manipulating photons in a single degenerate main cavity, where
photons can be adiabatically pumped to target OAM lattice sites ({i.e.}, the
OAM states). Thanks to the topological nature of the pumping, precise switch
can be realized without involving precise control of the parameters during
the pumping cycle. Such robustness of the pumping against perturbations as
well as its properties in the presence of photon losses are considered.
Moreover, even though the pumping process is adiabatic, the pumping cycle
can be very short. With realistic optical elements, it is possible to
achieve a switching rate of $\sim $MHz, which is much faster than commonly
used OAM switching devices \cite{thalhammer2013speeding,
mirhosseini2013rapid}. Even for a large change of OAM at $\sim 10^{q}$, only
$q$ degenerate main cavities and at most $5q$ pumping cycles are needed
using a multistage setup, leading to exponential speedup of the OAM switch.
The proposed topological photonic OAM switch only relies on a simple optical
setup with a few degenerate main cavities, therefore it may become a
powerful device for broad applications in quantum information and optical
communications. Our proposed topological approach for device engineering may
go beyond the OAM switch and inspire novel topological design of
conventional optical devices.

\begin{figure}[t]
\includegraphics[width=1.0\linewidth]{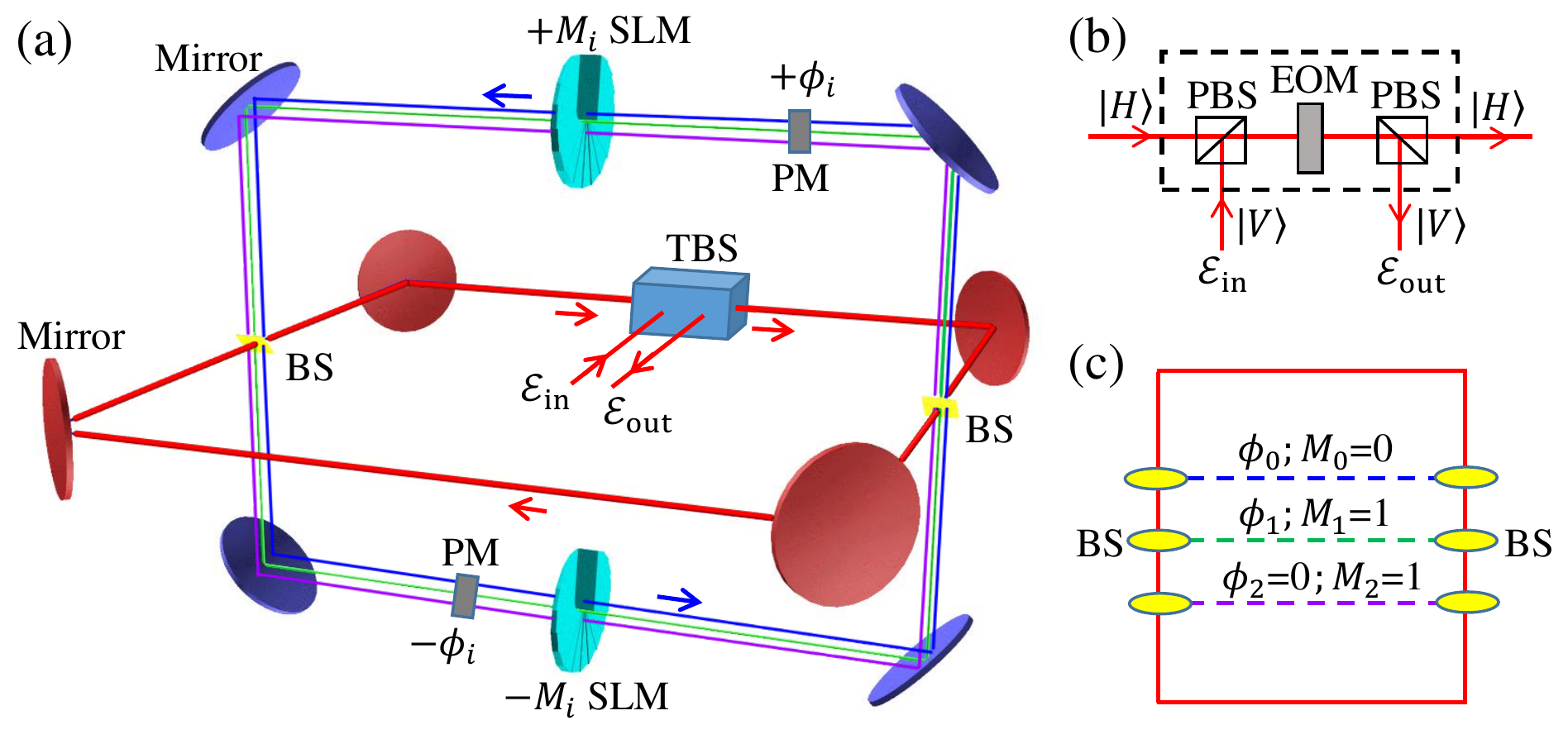}
\caption{(a) Experimental photonic circuit for a degenerate cavity system.
The main cavity (red curved mirrors) is coupled with auxiliary cavities
(blue curved mirrors) by beam splitters (BSs). A tunable beam splitter (TBS)
is used to tune the coupling between the cavity (in $H$ polarization) and
the input/output fields (in $V$ polarization). (b) Optical design for the
TBS. (c) Schematic diagram of the system with three auxiliary cavities
(dashed lines), with corresponding tunneling phase $\protect\phi _{s}$ and
step $M_{s}$ as labeled.}
\label{fig:system}
\end{figure}

\section{The system}
As shown in Fig.~\ref{fig:system}(a), our system
contains a main degenerate multimode cavity~\cite{arnaud1969degenerate,
horak2002optical, gopalakrishnan2009emergent, ballantine2017meissner,
schine2016synthetic}, which supports a large number of OAM modes and is
coupled with three degenerate auxiliary cavities by beam splitters.
The degenerate cavities used in our scheme possess fully transverse degeneracy
and their design principle is based on the round-trip Gauss matrix [often called
$ABCD$ ray matrix with $A,D$ ($B,C$) diagonal (off-diagonal) matrix
elements]~\cite{arnaud1969degenerate}. We consider a ring-type cavity possessing cylindrical symmetry with
respect to the optical axis, the cavity modes are Laguerre-Gaussian
(LG) modes $E_{p,l}$ with radial index $p$ and azimuthal index $l$. The
resonance frequencies of the LG modes are~\cite{yariv2007photonics,
hodgson2005laser}
\begin{equation}
\omega _{p,l}=n\Omega _{\text{F}}+(2p+|l|+1)\frac{\arccos (\frac{A+D}{2})}{%
2\pi }\Omega _{\text{F}},
\end{equation}%
where the integer $n$ is the longitudinal mode index that is fixed in our
scheme and $\Omega _{\text{F}}$ is the free spectral range of the cavity.
The off-diagonal ray matrix elements $B$ and $C$ only affect beam waist
size. The degenerate cavity is obtained when the round-trip $ABCD$ ray
matrix is equal to identity~\cite{arnaud1969degenerate}, which can be
achieved using curved mirrors or intracavity lenses. In this case, all
transverse LG modes $E_{p,l}$ have the same resonance frequency $n\Omega _{%
\text{F}}$, yielding a degenerate cavity supporting different OAM modes $l$
for photons. The tunneling between
different OAM states is realized by spatial light modulators (SLMs).
The unimportant radial index $p$, which is related to the radial
mode profile of the input photons, is omitted.
There are many ways to construct a degenerate cavity,
for example, we consider a
rectangular-shaped cavity shown in Fig.~\ref{fig:system}(a)
formed by four identical curved mirrors with focal length equal
to optical path length between adjacent mirrors (see Appendix A).
We use a tunable beam splitter (TBS) to couple the cavity with the
outside world, it is realized
by sandwiching an electro-optic modulator (EOM)~\cite{yariv2007photonics} 
between two polarizing beam splitters (PBSs), as shown
in Fig.~\ref{fig:system}(b). The EOM
rotates photon's polarizations in a
tunable way as $|H\rangle \rightarrow \sqrt{1-r_{\text{P}}^{2}}|H\rangle +r_{%
\text{P}}|V\rangle $ and $|V\rangle \rightarrow \sqrt{1-r_{\text{P}}^{2}}%
|V\rangle -r_{\text{P}}|H\rangle $, with $|H\rangle $ and $|V\rangle $ the
horizontal and vertical polarization states which are separated by the PBS.
The tunable coefficient $r_{\text{P}}$ acts as the reflectivity of the TBS.

The beam
splitters divert a small portion of the main-cavity photons towards the $s$%
-th auxiliary cavity, and merge them back after passing through spatial
light modulators (SLMs)~\cite{oemrawsingh2005experimental,
karimi2009efficient}, which change the OAM state by $\pm M_{s}$. This
corresponds to tunnelings between different OAM states in the main cavity,
with a tunneling rate determined by the reflectivities of the BSs. During
the tunneling, the photon can also acquire a phase determined by the optical
path-length difference between two arms of the auxiliary-cavity, which can
be generated and tuned using high-speed phase modulators (PMs). The system
is equivalent to a one-dimensional lattice model with the lattice sites
represented by the OAM states~\cite{luo2015quantum, zhou2017dynamically, luo2017synthetic}. The Hamiltonian is
given by (see Appendix A)
\begin{equation}
H=-\sum\nolimits_{l}\sum\nolimits_{s=0}^{2}J_{s}e^{i\phi
_{s}}a_{l+M_{s}}^{\dagger }a_{l}+h.c.,  \label{eq:H2}
\end{equation}%
where $a_{l}$ is the annihilation operator of the cavity photon in the OAM
state $l$, and $M_{s}$ is the step index of the SLMs in the $s$-th auxiliary
cavity with the tunneling amplitude $J_{s}$ and phase $\phi _{s}=l\alpha
_{s}+\beta _{s}$. The PM contains a beam rotator~\cite{leach2002measuring}
and an EOM~\cite{yariv2007photonics}: the beam
rotator, realized by two Dove prisms rotated by $\alpha _{s}/2$ with respect
to each other, is used to generate the $l$-dependent phase $l\alpha _{s}$,
and the EOM is used to tune the $l$-independent phase $\beta _{s}$.

The auxiliary cavities are designed as $\phi _{2}=0$, $M_{0}=0$, $%
M_{1}=M_{2}=1$, $J_{1}=J_{2}$, as shown in Fig.~\ref{fig:system}(c).
Therefore the Hamiltonian describes a generalized AAH model~\cite%
{ganeshan2015constructing} with modulations in on-site energy, tunneling
phase and amplitude, whose periods are determined by $\alpha _{0}$ and $%
\alpha _{1}$. Here, we focus our discussion on $\alpha _{0}=\alpha _{1}=2\pi
\cdot \frac{1}{2}$, in analogy to the Rice-Mele model (i.e., double-well
supper-lattice model)~\cite{rice1982elementary} with neighboring site energy
detuning $\Delta \equiv -4J_{0}\cos (\beta _{0})$ and inter- and intra-cell
tunnelings $J_{\pm }\equiv J_{1}[1\pm e^{i\beta _{1}}]$ [see Fig.~\ref%
{fig:loop}(a)].
We rewrite the Hamiltonian as
\begin{eqnarray}
H&=&-\sum\nolimits_{j}J_{+} b_{j,2}^{\dagger }b_{j,1}+J_{-}
b_{j+1,1}^{\dagger }b_{j,2}+h.c.  \nonumber \\
& &+\sum\nolimits_{j}\frac{\Delta}{2}(b_{j,1}^{\dagger
}b_{j,1}-b_{j,2}^{\dagger }b_{j,2}),  \label{eq:Ha1}
\end{eqnarray}
where we have introduced the unit cell index $j$ with $b_{j,1}=a_{2j}$ and $%
b_{j,2}=a_{2j+1}$. After a Fourier transformation $b_{k,1(2)}\propto\sum_j
e^{ijk}b_{j,1(2)}$, the Hamiltonian in the Bloch basis becomes
\begin{eqnarray}
H&=&-\sum\nolimits_{k} [b^\dag_{k,1},b^\dag_{k,2}]H_k[b_{k,1},b_{k,2}]^T,
\label{eq:Ha2}
\end{eqnarray}
with the Bloch Hamiltonian
$H_k(t)=\mathbf{h}%
(k,t)\cdot \vec{\sigma}$, where $\vec{\sigma}=(\sigma_x,\sigma_y,\sigma_z)$ are the Pauli matrix
and the real vector $\mathbf{h}(k,t)$ satisfies $h_{x}(k,t)+ih_{y}(k,t)=-[J_{+}(t)+J_{-}^{%
\ast }(t)e^{-ik}]$, and $h_{z}(k,t)=\frac{\Delta (t)}{2}$.
The band structure and the Bloch wave function $e^{ikj}|u_n(k)%
\rangle $
can be obtained by solving $H_k|u_n(k)\rangle=E_n|u_n(k)%
\rangle$, with band spectrum $E_{\pm}=\pm |\mathbf{h}(k,t)|$.
The parameters $\Delta=-4J_0\cos(\beta_0)$ and $J_{\pm}=J_1(1\pm
e^{i\beta_1})$ depend on the time-varying phases $\beta_{0,1}(t)$ [see Fig.~\ref{fig:loop}(b)],
so does the Hamiltonian $H(t)$.

\section{Topological pumping}
Topological pumping was first proposed by Thouless
for fermionic systems~\cite{thouless1983quantization}.
Consider a fermionic system with the same single-particle
Hamiltonian as $H(t)$,
if the Hamiltonian is modulated adiabatically and
periodically without closing the band gap, the amount of transported particles (along the OAM space) for a filled band is
characterized by the Chern number defined as the change in polarization during one pump cycle (i.e., the
center-of-mass displacement of the Wannier function), which is
\begin{equation}
\mathcal{C}_n=\frac{1}{2\pi}\int_0^Tdt\int_0^{2\pi} dk \Omega_{kt}
\end{equation}
with
\[
\Omega_{kt}=\langle \partial_t u_n(k,t)|\partial_k u_n(k,t)\rangle-\langle
\partial_k u_n(k,t)|\partial_t u_n(k,t)\rangle.
\]
The Chern number also
equals to the winding number of $\pm \mathbf{h}(k,t)$ surrounding the origin
as $t$ varies over one period and $k$ varies over the Brillouin zone.
The two band is gapped in the parameter space except the critical point when
$\Delta=|J_{+}|-|J_{-}|=0$. It can be proven that a loop which encloses the
critical point is topological non-trivial. For a clockwise loop enclosed the
critical point, we find that $\mathcal{C_{\pm}}=\mp1$.
The Chern number is
equal to the winding number of $\pm\mathbf{h}(k,t)$ surrounding the origin
as $t $ varies over one period and $k$ varies over the Brillouin zone.

Photons are non-interacting bosons, so the photonic pumping is reduced to
single-particle pumping, which is
different from the fermionic pumping of
a filled band. We consider a single photon in the $n$-th band
\begin{equation}
|\Psi(j,0)\rangle=\sum_{k}\psi_{k}e^{ijk}|u_n(k,0)\rangle,
\end{equation}
the modulation is adiabatic so that
it will follow the
initial band, and also $k$ is a good quantum number during the pumping.
So, the final state can be written as
\begin{equation}
|\Psi(j,t)\rangle=\sum_{k}\psi_{k}e^{ijk}e^{-i\int_0^t dt^{\prime
}E_n(k,t^{\prime })}e^{i\gamma_n(k,t)}|u_n(k,t)\rangle,
\end{equation}
with $\gamma_n(t)$ the Berry phase given by
\begin{equation}
\gamma_n(k,t)=i\int_0^t \langle u_n(k,t^{\prime })|\partial_{t^{\prime }}
u_n(k,t^{\prime })\rangle dt^{\prime }.
\end{equation}
The center of mass of the particle is
\begin{eqnarray}
\bar{j}(t)&=&\sum\nolimits_j\langle \Psi(j,t)|j|\Psi(j,t)\rangle \nonumber \\
&=&\sum_k\langle
\widetilde{\Psi}(k,t)|i\partial_k|\widetilde{\Psi}(k,t)\rangle
\end{eqnarray}
with
\begin{equation}
|\widetilde{\Psi}(k,t)\rangle=\psi_{k}e^{-i\int_0^t dt^{\prime
}E_n(k,t^{\prime })}e^{i\gamma_n(k,t)}|u_n(k,t)\rangle.
\end{equation}
So we have
\begin{equation}
\bar{j}(t)=\sum\nolimits_k i\psi^*_k\partial_k\psi_k+\int_0^tI(t^{\prime })
dt^{\prime }
\end{equation}
with the average current given by
\begin{eqnarray}
I(t)=\sum_k|\psi_k|^2[\partial_k E(k,t) + \Omega_{kt}].
\end{eqnarray}
So the displacement after one pumping cycle would be
\begin{eqnarray}
\Delta\bar{j}&=&\bar{j}(T)-\bar{j}(0)  \nonumber \\
&=&\int_0^TI(t) dt.
\end{eqnarray}

For the simple case with $\psi_k=\frac{1}{\sqrt{N}}$, $N$ is the total
number of lattice sites, $|\Psi(j,0)\rangle$ is reduced to the Wannier
function $|W_n(j,0)\rangle$,
\begin{equation}
|\Psi(j,0)\rangle=\sum_{k}\frac{1}{\sqrt{N}}e^{ijk}|u_n(k,0)\rangle%
\equiv|W_n(j,0)\rangle,
\end{equation}
and we have
\begin{eqnarray}
\Delta\bar{j} = \frac{1}{N}\sum_k\int_0^T
\Omega_{kt}dt=\int_0^Tdt\int_0^{2\pi}\frac{dk}{2\pi} \Omega_{kt}=\mathcal{C}%
_n.
\end{eqnarray}
We can see that the average displacement is exactly quantized even for a
single-particle Wannier state.

\begin{figure}[tbp]
\includegraphics[width=1.0\linewidth]{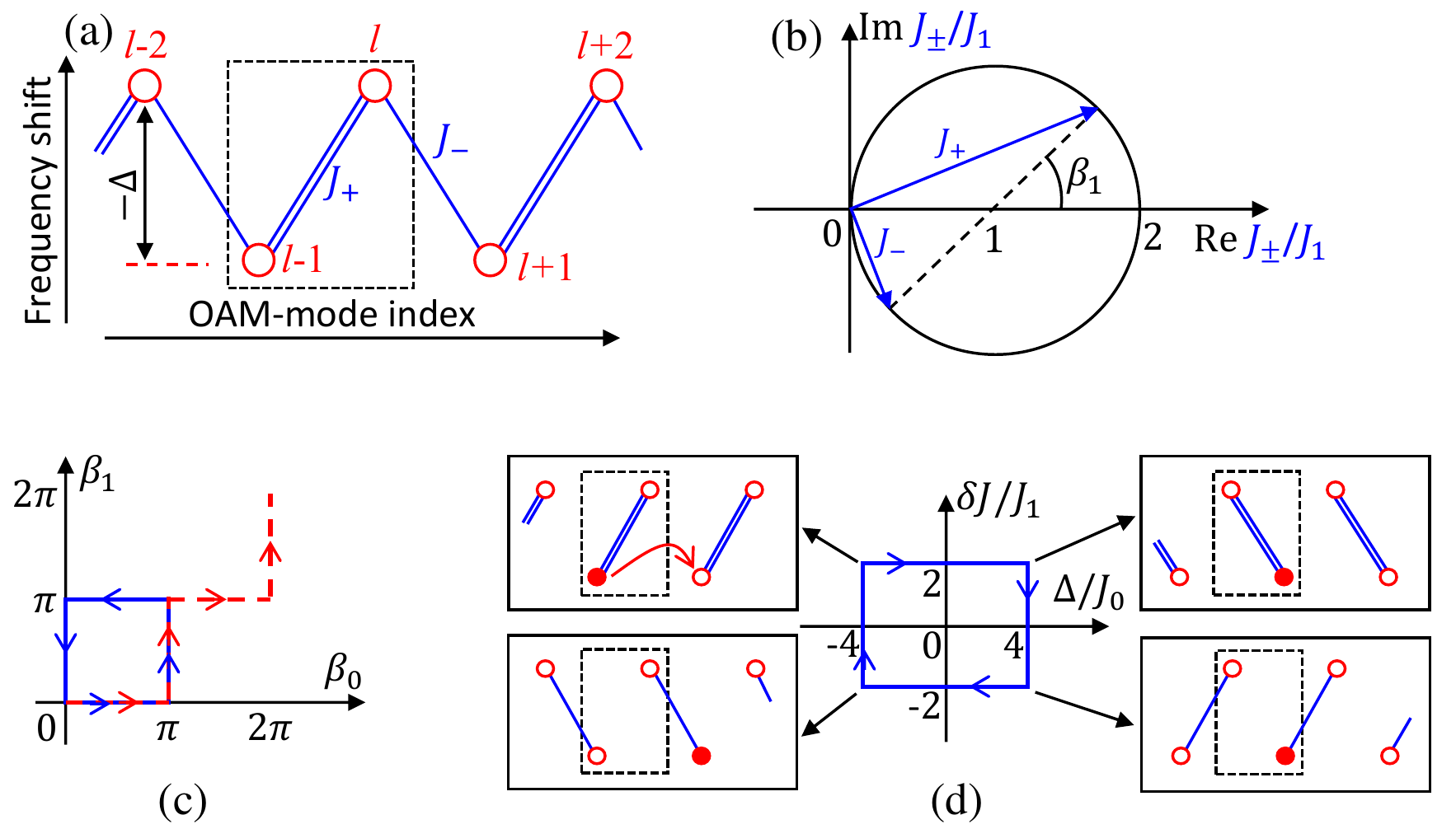}
\caption{ (a) Diagram of the effective lattice in the OAM dimension. Each
unit cell (enclosed by the dashed square) contains two sites with detuning $%
\Delta =-4J_{0}\cos [\protect\beta _{0}(t)]$. $J_{\pm }=J_{1}[1\pm e^{i%
\protect\beta _{1}(t)}]$ are the inter- and intra-cell tunnelings
respectively, whose dependence on the tunneling phase $\protect\beta _{1}$
is shown in (b). (c) The pumping loop in the $\protect\beta_{0}$-$\protect%
\beta_{1}$ plane. Since the tunneling phase has a period of $2\protect\pi $,
their choice is not unique, and the two loops (red dashed and blue solid)
have the same topology. (d) Illustration of the pumping process. The loop in
the $\protect\delta J$-$\Delta $ plane can be realized by either loop shown
in (c). Red solid circle represents a site occupied by a photon, which moves
to the right by two sites (one unit cell) during one pumping cycle.}
\label{fig:loop}
\end{figure}

\begin{figure}[tbp]
\includegraphics[width=1.0\linewidth]{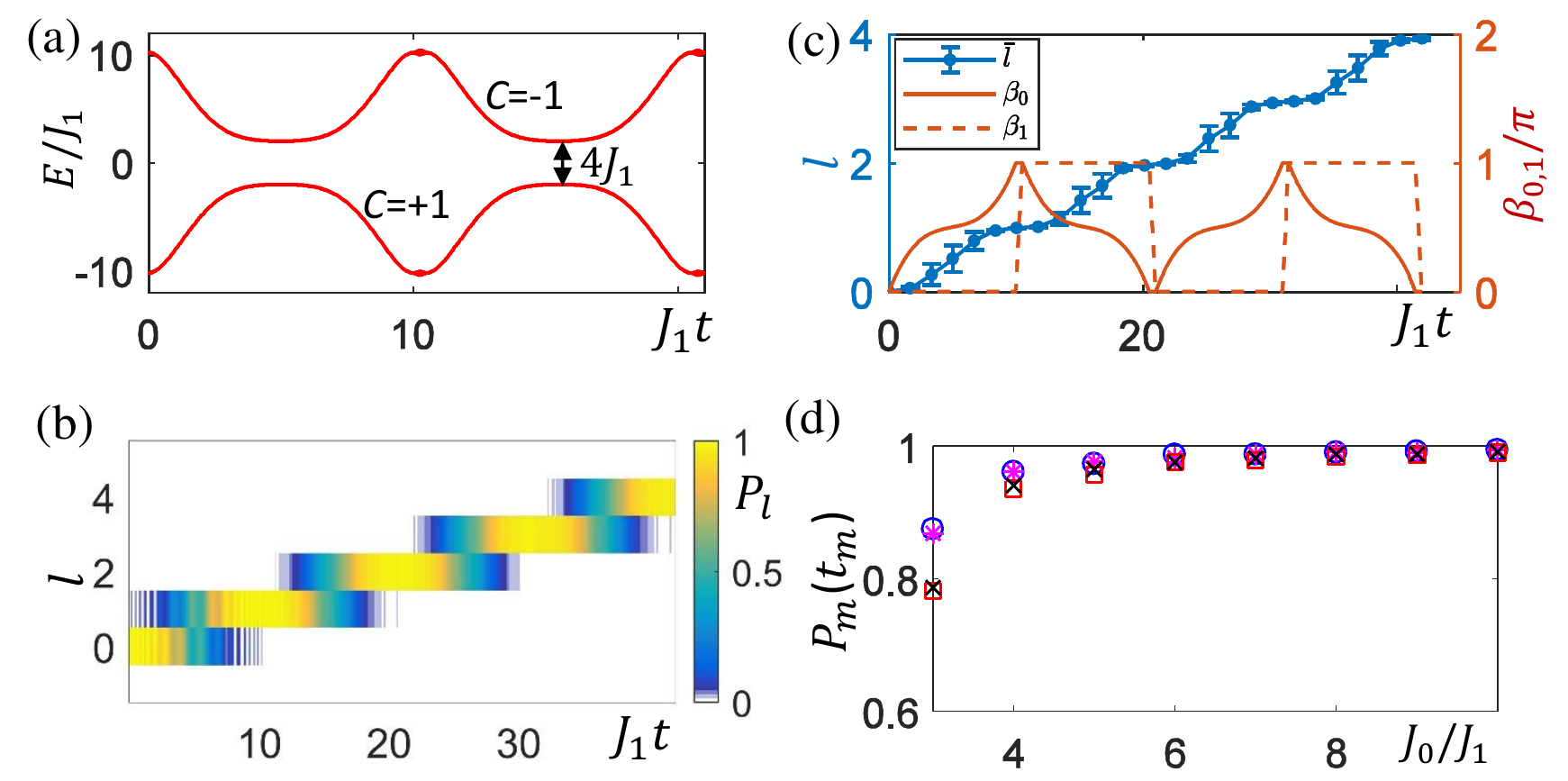}
\caption{(a) The band structure in one pump cycle with Chern number $%
\mathcal{C}=\pm 1$ for two bands. (b) Cavity photon distribution in two pump
cycles with step-like transport. $P_{l}(t)$ is the probability of the photon
in OAM state $l$ at time $t$. (c) Center of mass displacements (dots) and
their standard deviations (error bars), which are calculated by assuming
Gaussian distributed random disorders $\protect\delta \protect\beta _{0,1}$
in the tunneling phases with standard deviations $\protect\sigma (\protect%
\delta \protect\beta _{0,1})=0.1$rad. (d) The final-state purities at $\frac{%
m}{2}$ pump cycles ($t_{m}=\frac{mT}{2}$) versus different coupling ratio $%
J_{0}/J_{1}$. The pink stars, blue cycles, black crosses and red squares
correspond to $m=1$, $2$, $3$ and $4$ respectively. The modulation of the
pump parameters is chosen to satisfy the adiabatic condition with period $%
T=21/J_{1}$, whose temporal profiles are shown in (c) with $\protect\beta %
_{0}$ the red solid line and $\protect\beta _{1}$ the red dashed line. In
(b-d) the initial OAM state is $l_{0}=0$ and in (a-c) the coupling ratio is $%
J_{0}/J_{1}=5$.}
\label{fig:displacement}
\end{figure}

We consider a pumping process as shown in Fig.~\ref{fig:loop}(c), where the pumping cycle
corresponds to a loop in the 2D parameter space spanned by $\beta _{0}$ and $%
\beta _{1}$. The two bands of the system are gapped in the parameter space
except at the critical point $\beta _{0}=\beta _{1}=\pi /2$ with $\Delta
=\delta J=0$ and $\delta J\equiv |J_{+}|-|J_{-}|$. A loop which encloses
this critical point is topologically non-trivial, and the topology of the
pump is invariant under deformation of the loop without cutting through the
critical point. Therefore it is more convenient to consider the pump in the $%
\Delta $-$\delta J$ plane, with corresponding pump loop and photon movement
illustrated in Fig.~\ref{fig:loop}(d).

The band structures ($E_{\pm }=\pm |\mathbf{h}(k,t)|$) along the pumping
loop are shown in Fig.~\ref{fig:displacement}(a), with a smallest bandgap $%
4J_{1}$. The two gapped bands have different transport properties due to
their different topologies, characterized by Chern number $\mathcal{C}=\mp 1$
respectively. We start the pumping at $\delta J=2J_{1}$ and $%
\Delta =-4J_{0}$ with $J_{0}\gg J_{1}$, where the Wannier functions are well
localized at a single OAM state (since the inter-cell tunneling $J_{-}$ is
vanished and intra-cell tunneling $J_{+}$ is very weak compared to detuning $%
\Delta $). We consider a lower-band state with a single photon initialized
at OAM state $l_{0}$ (this can be realized by resonantly feeding a
single-photon pulse carrying the corresponding OAM into the cavity), which
will be pumped to $l_{0}+2$ after one cycle (each unit cell contains 2
sites). The photon distribution and displacement of its center-of-mass are
shown in Figs.~\ref{fig:displacement}(b) and (c). We can see a clear
step-like displacement as expected. The transport is topologically protected
and robust against perturbations in the parameter modulation loop. Fig.~\ref%
{fig:displacement}(c) shows the center-of-mass displacement and its
variation caused by the random shifts in the phase $\beta _{0}$ and $\beta
_{1}$, demonstrating that the quantized transport is almost immune to such
small errors.

\begin{figure*}[!htbp]
\includegraphics[width=0.8\linewidth]{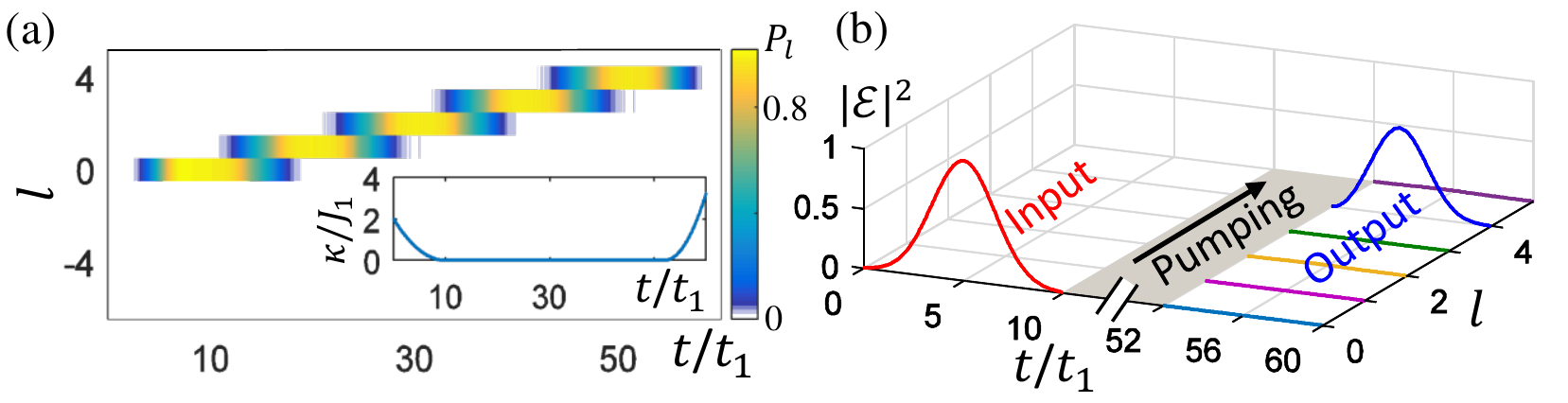}
\caption{ (a) Normalized cavity photon distribution and (b) Input/output
photon pulses during the OAM switching. The normalized input pulse is $%
\mathcal{E}_{\text{in}}(t)=e^{-iE_{-}t-0.2(J_{1}t-5)^{2}}$ with $l_{0}=0$
(i.e., only the zero-OAM Wannier orbital is excited). $E_{-}$ is the lower
band energy. After the input pulse enters the cavity, we pump the state to
higher OAM modes, and then release the signal by increasing the cavity loss $%
\protect\kappa $ [see the inset in (a)]. Other parameters are the same as in
Fig.~\protect\ref{fig:displacement}(b).}
\label{fig:switch}
\end{figure*}

To obtain a final state with a high purity (defined as the probability of
finding the photon in the desired OAM state), a large $J_{0}/J_{1}$ is
required to make the Wannier function well localized at a single OAM state.
This is because,
at the beginning (end) of the pumping, we have two flat
bands with $E_{\pm}=\pm\sqrt{4J_0^2+4J_1^2}$, and the eigenvectors $%
|u_-(k,0)\rangle=[\cos(\theta),-\sin(\theta)]$ $|u_+(k,0)\rangle=[\sin(%
\theta),\cos(\theta)]$, with $\tan(2\theta)=J_1/J_0$ independent of $k$. The
Wannier functions of the lower band is $|W_{-}(0)\rangle=\cos(\theta)|l=0%
\rangle-\sin(\theta)|l=1\rangle$, which is well localized on $l=0$ for $%
J_1/J_0\ll1$. As a result, the pumping of state $|l=0\rangle$ is
characterized by the lower band which gives a quantized transport of $%
\mathcal{C}_{-}=1$.
In addition, different from the pumping of a whole filled band,
the final state is not a simple displacement
of the initial state for single-particle
pumping, this is because there exists diffusion during the pumping
due to the dynamical phase $e^{-i\int_0^T dt E_n(k,t)}$,
which decreases the purity of the final state.
The diffusion leads to a wider profile of its density
distribution, and induces minor populations of unwanted OAM states.
To reduce such effect, we need to make the band as flat (in
momentum space) as possible during the
pump, then the dynamical phase becomes a constant phase
independent of $k$. During the pump,
the bands are given by
\begin{eqnarray}
E_{\pm}&=&\pm\sqrt{4J_0^2+|J^*_{+}+J_{-}e^{ik}|^2}   \\
&\simeq&\pm\left\{2J_0+\frac{J_1^2}{J_0}+\frac{J_1^2}{4J_0}%
[\cos(k)-\cos(k+2\beta_1)]\right\},  \nonumber
\end{eqnarray}
which is
always flat for $\beta_1=0,\pi$. As we modulate $%
\beta_1$ from $0$ ($\pi$) to $\pi$ (0), such that $J_+$ ($J_-$) changes from $2J_1$ to $0$ while
$J_-$ ($J_+$) changes from $0$ to $2J_1$, the bands are
approximately flat in the limit $J_1/J_0\ll1$, and the
diffusion effect is negligible. The band gap is also very large during this
modulation, and the diffusion effect can be reduced further by increasing
the modulating speed properly.
These effects are verified by our
numerical calculation of the purity for different values of $J_{0}/J_{1}$
[see Fig.~\ref{fig:displacement}(d)].
The purity can be close to 100\% by using a larger value of $J_{0}/J_{1}$.

Each pumping cycle shifts photon's OAM by 2, therefore even switching
numbers can be realized by integer pumping cycles. However, we notice that
photon's OAM state is also well localized at half-integer pumping cycles
with OAM shifted by an odd number [see Figs.~\ref{fig:displacement}(b)-(d)],
therefore odd switching numbers can be realized by half-integer pumping
cycles. Alternatively, we can design the synthetic double-well lattice such
that two sites in each unit cell are represented by different polarization
states with the same OAM, and each pumping cycle changes photon's OAM by 1.
Similarly, we can put the photon in the upper band,
then the above pumping will change
its OAM by $-1$ or $-2$, alternatively, this can also be
achieved by considering
a counterclockwise pumping loop and a photon in the lower band.
The system is linear (with no photon-photon interaction), as a result,
our scheme can simultaneously switch multiple photons in different OAM modes, each pumping
cycle shifts the OAM by $+2$ ($-2$) for photons in the lower (upper) band.

\section{Topological OAM switch}
Such quantized transport offers a robust way to switch the OAM states of
photonic signals in three steps: (i) Input---the input photon pulse enters
the cavity and the $l_{0}$ cavity-mode is excited; (ii) Topological
pumping---the photon is pumped to the desired OAM state; (iii)
Output---photon pulse is released out of the cavity. The input and
output are realized by a tunable beam splitter (TBS) which couples the cavity with the
outside world. The tunability of the input/output beam splitter is crucial for improving the efficiency of the OAM
switch because the coupling between the cavity and the outside world need be
turned on during input/output so that the signals can get in/out, and turned
off during the pumping to avoid unwanted photon losses.
The dynamics are characterized by \cite{walls2008quantum}
\[
\dot{{a}}_{l}=\frac{1}{i}[a_{l},H]-\frac{\kappa }{2}{a}_{l}+\delta _{l,l_{0}}%
\sqrt{\kappa _{\text{e}}}\mathcal{E}_{\text{in}}(t),
\]%
where $\mathcal{E}_{\text{in}}(t)$ is the input photonic field in $l_{0}$%
-OAM state, $\kappa $ is the total photon loss, and the tunable coupling
strength between the cavity and the input/output fields is $\sqrt{\kappa _{%
\text{e}}}\simeq |r_{\text{P}}|\sqrt{\frac{\Omega _{\text{F}}}{2\pi }}$ \cite%
{hernandez1986fabry} with $r_{\text{P}}^{2}\ll 1$ and $\Omega _{\text{F}}$
the free spectral range (FSR) of the main cavity.

In an ideal case, all optical elements are perfectly designed, and the only
photon loss channel is the TBS, so we have $\kappa =\kappa _{\text{e}}$.
With proper modulation of $\kappa _{\text{e}}$ (i.e., $r_{\text{P}}$), the
input signal pulse can enter the cavity with an efficiency as high as $90\%$%
. Then it is pumped to the desired OAM state and finally released from the
cavity. The evolution of the photon field inside the cavity, as well as the
temporal profiles of input/output fields and photon loss $\kappa $ are shown
in Figs.~\ref{fig:switch}(a) and (b). We see that photons are almost
perfectly transported to the desired OAM state, but the intensity is
slightly reduced due to photon losses during input and diffusion during
pumping.

\begin{figure*}[!htbp]
\includegraphics[width=0.8\textwidth]{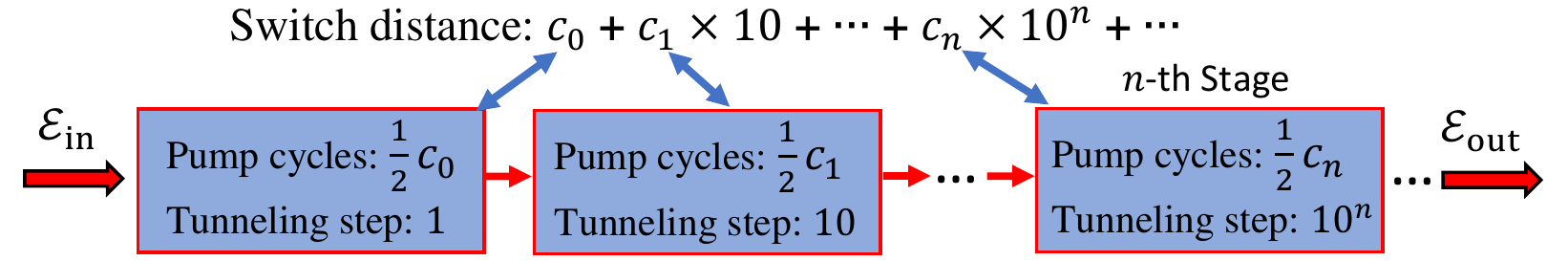}
\caption{Illustration of a multistage setup with $N=10$. Each half pump
cycle shifts the OAM by the corresponding tunneling step.}
\label{fig:multistage}
\end{figure*}

\section{Experimental consideration}
Typically the FSR of the main cavity is
$\Omega _{\text{F}}\sim 2\pi \times 1$GHz, and the tunneling strength, given
by $J_{s}=\Omega _{\text{F}}\frac{|r_{s}|^{2}}{2\pi (1+|t_{s}|^{2})}$ with $%
r_{s}$ ($t_{s}$) the reflectivity (transmissivity) of the corresponding beam
splitters \cite{hafezi2011robust, luo2015quantum, zhou2017dynamically, luo2017synthetic}, can be up to tens of MHz
(e.g., $J_{0}\sim 2\pi \times 20$MHz and $J_{0}\gg J_{1}\sim 2\pi \times 4$%
MHz). The pump period $T\sim 20/J_{1}$ leads to a switching time of the
order of $\mu $s.
Notice that the bandwidth of the input signal pulse should be smaller than
the initial band gap $4J_0$, leading to a bandwidth
of the order of $10$MHz for the input photon, and such photon source can be
realized using cavity-enhanced parametric downconversion~\cite{PhysRevLett.101.190501}.
The upper limit ($l_{\text{max}}$) that the OAM state can be switched to, is
determined by the aperture of the optical elements because the beam size
increases with the OAM number $l$. For typical mirror size, $l_{\text{max}}$
can be very large (hundreds), leading to a switchable OAM range $l\in
\lbrack -l_{\text{max}},l_{\text{max}}]$. Imperfections such as extra photon
losses may be introduced by cavity mirrors, TBS, phase modulators and SLMs. Such
losses only reduce the intensity of the output field without affecting the
quantized switching (see Appendix B). To reduce such
extra losses, we can make use of high-efficiency intra-cavity elements (e.g., EOM, SLM, etc.)
with high transmission (through anti-reflection coating),
it is possible to make these extra losses as small as a few tens of kHz
(much smaller than the switching rate of MHz)~\cite{nagorny2003collective,
oemrawsingh2005experimental, marrucci2011spin, raut2011anti}.
Also, the defects in the SLMs would induce imperfect switch,
fortunately, only two high-efficiency SLMs
(which can be fabricated with high precision~\cite{oemrawsingh2005experimental}) are used and
photons may pass the SLMs many times to realize high switch distance.
Deviations from cavity degeneracy lead to random on-site energy
shifts $\delta \omega _{l}\sim |l|\delta \omega $ (see Appendix C). Due to the
topological protection, our scheme works well as long as the shift for the
maximal OAM state (hundreds) is smaller than the smallest band gap $l_{\text{%
max}}\delta \omega \lesssim 4J_{1}$ (see Appendix C). This requires that the
accuracy of the mirror (lens) distance should be of the order of $\mu $m
which can be realized using high-precision translation stage.
Thanks to the
degenerate-cavity setup with all OAM modes sharing the same optical paths,
disorders in the tunneling coefficients are negligible since all tunnelings
are realized by the same sets of auxiliary cavities and can be controlled
simultaneously.
Other imperfections, such as a global shift in the resonance frequency of
the degenerate cavity, will not affect the pumping. Therefore, our scheme
should be feasible even in the presence of realistic imperfections.

\section{Multistage generalization}
To obtain high OAM states, large numbers
of pump cycles may be required, which slow down the switching rate. To
accelerate the switching rate, we consider a multistage setup with several
cascaded degenerate-cavity systems as shown in Fig.~\ref{fig:multistage}. In
the $n$-th stage, we choose the tunneling step of the SLMs as $%
M_{1}=M_{2}=N^{n}$ with $N$ an integer, and $\alpha _{0}=\alpha _{1} $
satisfying $\text{mod}(\alpha _{0}N^{n},2\pi )=\pi $. For an arbitrary
switching distance $\Delta l=\sum_{n}c_{n}N^{n}$ (with $c_{n}$ the expansion
coefficients), the corresponding pump cycle of the $n$-th stage is $c_{n}/2$%
, which gives the total switching time $T\sum_{n}c_{n}/2$. For example,
consider $\Delta l=512$ and $N=10$, the total switching time is only $4T$.
Both the maximum number of stages $\leq \log _{N}l_{\text{max}}$ and the
maximum switching time $\leq \frac{TN}{2}\log _{N}l_{\text{max}}$ are
logarithmic, yielding exponential speedup for the OAM switch.

\section{Discussion and Conclusion}
Conventional spatial light modulator
\cite{thalhammer2013speeding} and digital micro-mirror device \cite%
{mirhosseini2013rapid} are limited by the switching rates of $\sim $kHz.
Higher switching rates can be achieved by combing acousto-optic
(electro-optic) modulator with SLMs (q-plate) \cite{radwell2014high,
slussarenko2011efficient}, or using on-chip resonators \cite{strain2014fast}%
. However, the acousto-optic modulator would induce unwanted change in
wavelength, and the on-chip switching has a very low efficiency. Moreover,
all these approaches require precise control of experimental parameters and
the number of usable OAM modes is usually very limited. In contrast, our
scheme is robust against perturbations due to its topological feature, and
also able to rapidly switch to high OAM modes with high efficiency. Our
results of single-photon pumping can be generalized to multi-photon states
or even classic coherent states. Since the system is linear with no
interaction, every photon is pumped independently.

In summary, we proposed a topological photonic OAM switch which is fast,
robust, efficient and accessible to exponentially large OAM states. The
proposed topological pumping in the OAM-based synthetic dimension offers a
powerful platform to study 1D topological physics of the generalized AAH
model. The simple optical
setup for the topological photonic OAM switch opens a wide range of
experimental opportunities and may find important applications in quantum
information processing and optical communications.

\begin{acknowledgments}
\textbf{Acknowledgments}: This work is funded by NNSFC (Grant Nos. 11574294,
61490711), NKRDP (Grant Nos.2016YFA0301700 and 2016YFA0302700), the
\textquotedblright Strategic Priority Research Program(B)\textquotedblright\
of the CAS (Grant No.XDB01030200), ARO (W911NF-17-1-0128), AFOSR
(FA9550-16-1-0387), and NSF (PHY-1505496).
\end{acknowledgments}

\appendix
\setcounter{equation}{0} \renewcommand{\theequation}{A\arabic{equation}}

\section{Appendix A}
\begin{figure}[tbp]
\includegraphics[width=0.8\linewidth]{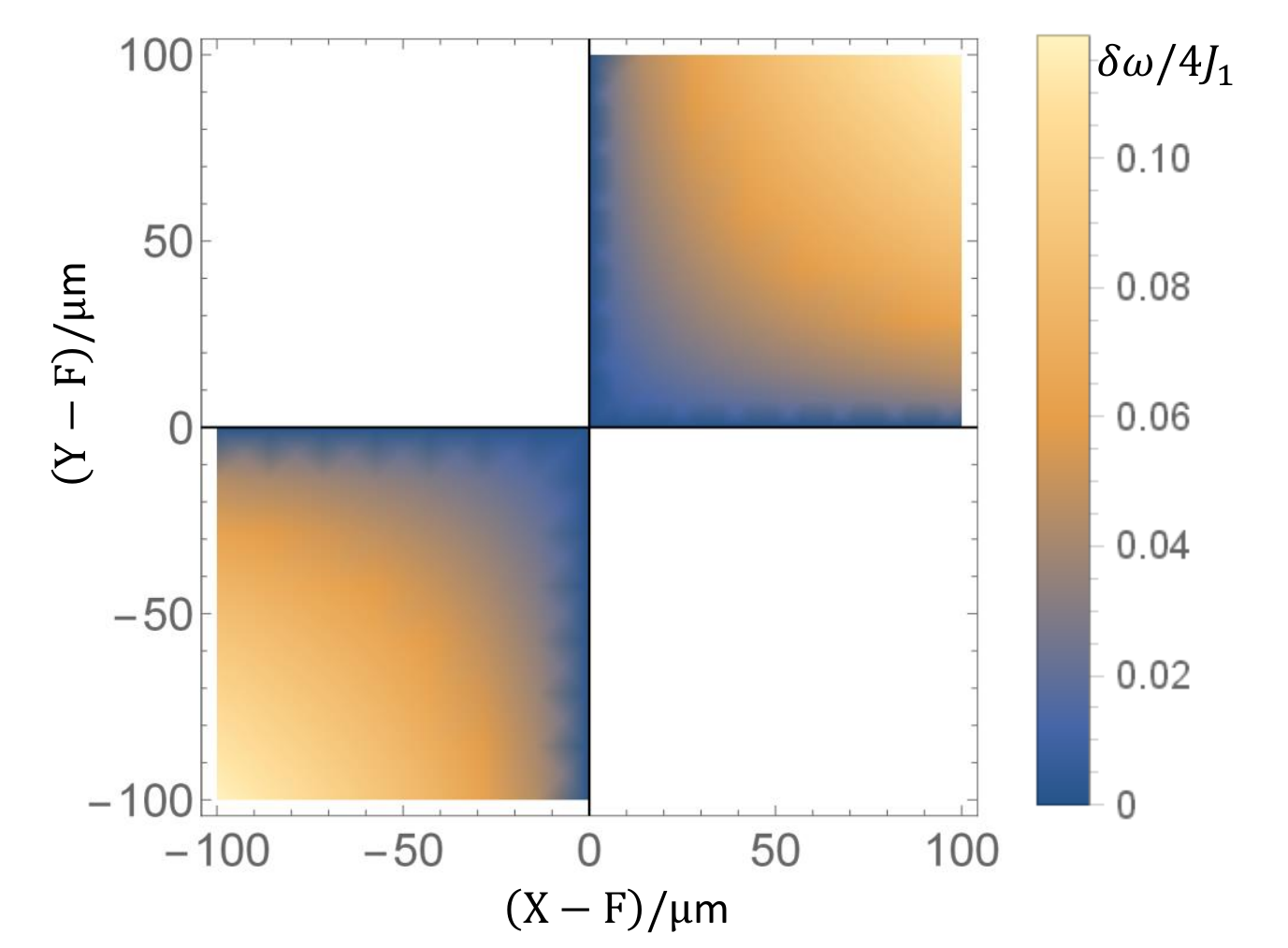}
\caption{Plot of frequency difference $\protect\delta \protect\omega $
between adjacent OAM modes normalized to the band gap $4J_{1}$ with respect
to the width and length of the rectangular cavity. The white regions
represent configurations where the cavity is unstable. Our scheme works well
if $|l|\frac{\protect\delta _{\protect\omega }}{4J_{1}}\lesssim 1$. Other
parameters are $F=10$cm, $J_{1}=0.004\Omega _{\text{F}}$.}
\label{fig:stable_a}
\end{figure}

\textbf{The cavity stability and effective Hamiltonian.}
There are many ways to construct a degenerate cavity,
and its stability may depend on specific configuration.
for example, we consider a
rectangular-shaped cavity shown in Figure 1(a) in the main text with width $%
X $ and length $Y$. Four curved mirrors are identical with the focal length $%
F$ and the fully transverse degeneracy can be obtained for $X=Y=F$. The
cavity is stable for $(X-F)(Y-F)>=0$ and unstable for $(X-F)(Y-F)<0$ (see
Fig.~\ref{fig:stable_a}). In the stable region, the beam waist is given by $%
w_{0}^{2}=\frac{\lambda F}{2\pi }\sqrt{\frac{X-F}{Y-F}}$ with $\lambda$ the wavelength,
and $w_{0}$ is of
the order of $1$mm. These properties are similar to those for confocal Fabry-Perot
cavity~\cite{yariv2007photonics, hodgson2005laser}. Because the cavity
possesses cylindrical symmetry with respect to the optical axis, ellipsoidal
(rather than spherical) mirrors are used to correct astigmatism caused by
non-normal reflections. Alternatively, this can be done by replacing curved
mirror with an intracavity lens and a plane mirror, or by changing the
geometry of the cavity.

Inserting beam splitters (BSs) may slightly adjust the optical path length
of the cavity and thereby the $ABCD$ ray matrix, which can be restored
easily to identity by modifying $X$ or $Y$. The beam splitters divert a
small portion of photons in the main cavity to the auxiliary cavity
containing two SLMs, which induce tunnelings between different OAM modes. To
show this, we first consider a single auxiliary cavity with two $\pm 1$%
-SLMs. Our system is equivalent to a 1D lattice of coupled cavity array with
lattice sites represented by different OAM modes~\cite{luo2015quantum, zhou2017dynamically, luo2017synthetic},
as shown in Fig.~\ref%
{fig:sys_a}. The optical path length of the auxiliary cavity is chosen for
destructive interference so that photons spend most of the time in the main
cavity (small imperfections of the auxiliary cavity can be ignored because
of this). The system is characterized by the following tight-binding
Hamiltonian~\cite{hafezi2011robust}
\begin{equation}
H=-\sum\nolimits_{l}Je^{i\phi }a_{l+1}^{\dagger }a_{l}+h.c.,  \label{eq:Ha}
\end{equation}%
where $a_{l}$ is the annihilation operator of the cavity photon in the OAM
state $l$, $J$ and $\phi $ are corresponding tunneling amplitude and phase,
which are determined by reflectivities of the BSs and the optical path
length of two arms of the auxiliary cavity, respectively. It is
straightforward to generalize the results to multi auxiliary cavities. Using
three properly designed auxiliary cavities, we obtain the Hamiltonian Eq.
(1) in the main text. Without tunnelings, the system has a single trivial
flat band since all OAM modes are degenerate. The tunneling and interference
between different OAM modes lead to non-trivial topological band structures,
based on which we can switch OAM states through topological pumping.

\begin{figure}[tbp]
\includegraphics[width=1.0\linewidth]{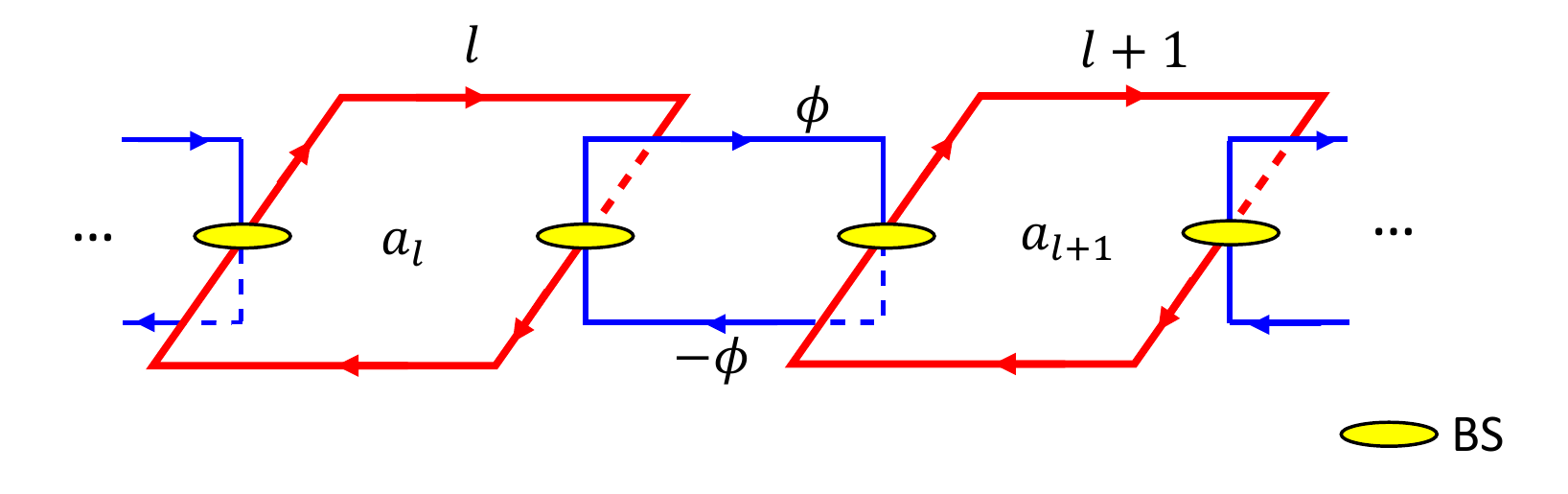}
\caption{The equivalent circuit in the OAM space, with $\protect\phi $ the
phase imbalance between two arms of the auxiliary cavity and $a_{l}$ the
field operator for OAM mode $l$. Red (blue) loop represents the main (auxiliary) cavity.}
\label{fig:sys_a}
\end{figure}

\begin{figure*}[tbp]
\includegraphics[width=0.8\linewidth]{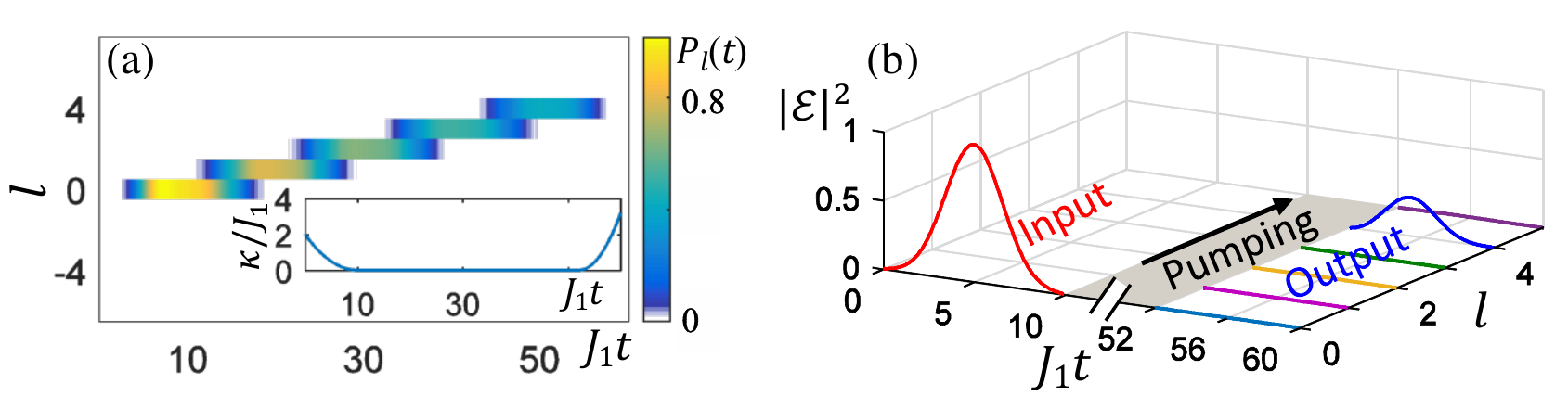}
\caption{ (a) Normalized cavity photon distribution and (b) Input/output
photon pulses in the OAM switching. The parameters are the same as in Fig.~4
in the main text, except that the photon loss is nonzero during the pumping
with $\protect\kappa_0=0.02J_1$.}
\label{fig:switch_a}
\end{figure*}

\section{Appendix B}
\textbf{Effects of photon losses}.
Typically, the optical path length of the cavity is about tens of
centimeters which leads to a FSR $\Omega_\text{F}=2\pi c/L\sim 2\pi\times1$%
GHz, where $L$ is the optical path length of the cavity and $c$ the speed of
light. The reflectivity $r_\text{P}$ of the TBS can be tuned by the EOM. If
the tuning rate is much smaller than the FSR $\Omega_\text{F}$, its effect
is well characterized by a time-dependent coupling $\sqrt{\kappa_\text{e}(t)}%
\simeq |r_\text{P}(t)|\sqrt{\frac{\Omega_\text{F}}{2\pi}}$, where $\kappa_%
\text{e}$ can be tuned from 0 to a few MHz within a few $\mu$s and vice
versa.
In realistic experiments, there are other photon losses due to factors such
as the finite Q-factor of the cavity, the intrinsic loss of the SLMs and
phase modulators. Such photon loss can be made as low as tens of kHz (much
smaller than the switching rate) using high performance optical elements.
For extremely high OAM states, the beam size becomes comparable with the
aperture of the optical elements, which gives a upper limit that the OAM can
be switched to. We consider the pumping between OAM states smaller than the
upper limit, the dynamics of the lossy system is characterized by the master
equation
\begin{eqnarray}
\dot{\rho}=\frac{1}{i}[H,\rho]+\kappa_0\sum_l (a_l\rho a_l^\dag -\frac{1}{2}%
\rho a_l^\dag a_l-\frac{1}{2}a_l^\dag a_l\rho)  \label{eq:master}
\end{eqnarray}
with photon loss rate $\kappa_0$. For the pumping of a single photon state,
the solution is simply given by
\begin{eqnarray}
\rho=(1-e^{-\kappa_0 t})|vac\rangle\langle vac|+e^{-\kappa_0
t}|\Psi(t)\rangle\langle \Psi(t)|,
\end{eqnarray}
with $|\Psi(t)\rangle$ the solution of non-dissipative case. The photon
density distribution
\begin{eqnarray}
N(l,t)&=&\frac{\text{Tr}[\rho(t)a_l^\dag a_l]}{\sum_l\text{Tr}%
[\rho(t)a_l^\dag a_l]}  \nonumber \\
&=&\langle \Psi(t)|a_l^\dag a_l|\Psi(t)\rangle
\end{eqnarray}
is the same as the non-dissipative case, except that the probability to find
the photon inside the cavity is reduced to $e^{-\kappa_0 t_\text{p}}$ with $%
t_\text{p}$ the total pumping time. Our system is linear with no
interactions, thus the results of single photon pumping can also be
generalized to multi-photon states and even classic coherent states.

When the cavity is coupled with the input/output signals, the dynamics are
characterized by the Langevin equation
\begin{eqnarray}
\dot{{a}}_l= \frac{1}{i} [a_l, H] -\frac{\kappa}{2}{a}_l + \delta_{l,l_0}%
\sqrt{\kappa_\text{e}}\mathcal{E}_{\text{in}}(t),  \label{eq:coher}
\end{eqnarray}
with $\kappa=\kappa_0+\kappa_\text{e}$ and $\mathcal{E}_{\text{in}}(t)$ the
input optical field in $l_0$-OAM state which can be either a single-photon
pulse or a classic coherent pulse. The dynamics of both single-photon and
coherent input signal are described by Eq.~(\ref{eq:coher}), with $a_l$
being the coherent (single-photon) amplitude of OAM state $l$. Typically,
the intrinsic photon loss $\kappa_0$ is of the order of tens of kHz (which
can be made even smaller by improving the performance of the optical
elements), and it only reduces the switching efficiency without affecting
the quantized transport (even for a strong loss), as shown in Fig.~\ref%
{fig:switch_a} with a large $\kappa_0$ ($\sim 2\pi \times 100$kHz).

\begin{figure*}[tbp]
\includegraphics[width=0.8\linewidth]{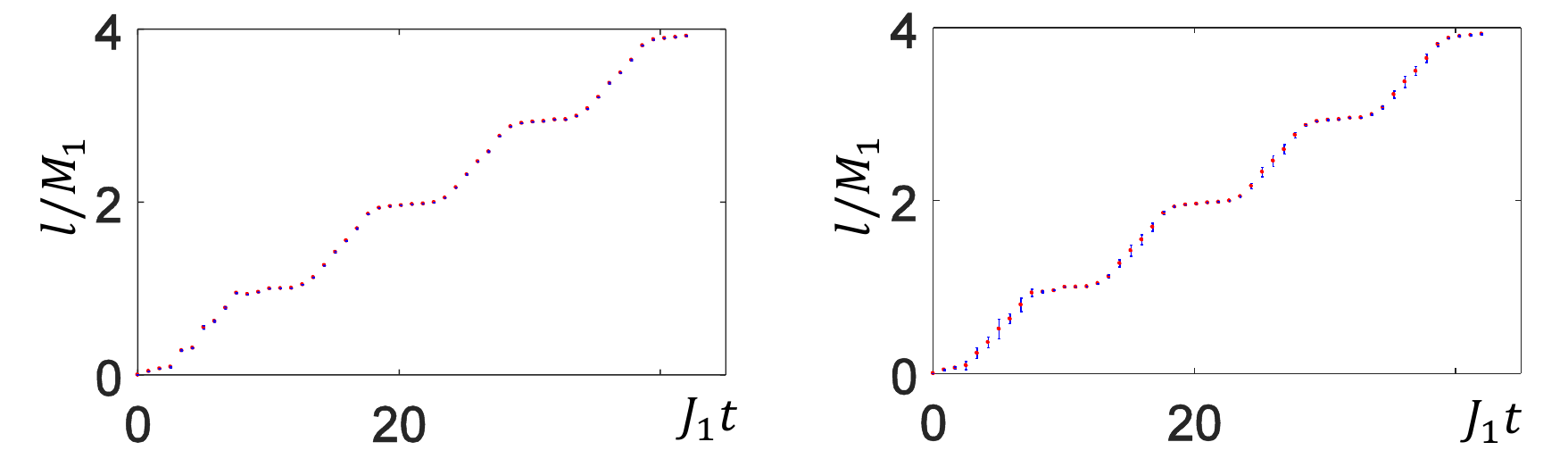}
\caption{(a) Center of mass displacements (red dots) and their standard
deviations (blue error bars), which are calculated by assuming Gaussian
distributed random on-site energy shift $\protect\delta \protect\omega %
_{l}=|l|\protect\delta \protect\omega $ due to deviations from the
degeneracy. Standard deviations $\protect\sigma (\protect\delta \protect%
\omega )=0.05J_{1}$ and tunneling step $M_{1}=1$ corresponds to the $0$-th
stage of the multi-stage switch. $\protect\delta \protect\omega =0.05J_{1}$
corresponds to $\protect\sqrt{(X-F)(Y-F)}\simeq 10\protect\mu $m. (b) The
same as in (a) except that the tunneling step $M_{1}=10$, corresponding to
the $1$-th stage of the multi-stage switch.}
\label{fig:error_a}
\end{figure*}

\section{Appendix C}
\textbf{Effects of deviations from degeneracy point.}
In the ideal case, the cavities always stay at the degeneracy point when $ABCD$ matrix is equal
to identity. Deviations from the degeneracy point lead to an frequency
shift $(2p+|l|+1)\Omega _{\text{F}}\arccos (\frac{A+D}{2})$ to the LG cavity modes.
This would hardly affect tunnelings because the BS reflectivity is not affected
and the optical path length of the auxiliary cavity is chosen for
destructive interference (photons spend most of their time in the main
cavity).
For the OAM states considered in our system, the radial index $p$ is of the
same order of magnitude as $|l|$ (though each OAM mode may contain multiple $%
p$ components). Overall, deviations from degeneracy lead to a random energy
shift $\delta \omega _{l}a_{l}^{\dagger }a_{l}$ to each OAM mode, with $%
\delta \omega _{l}\simeq |l|\delta \omega $ and $\delta \omega \simeq 3\frac{%
\arccos (\frac{A+D}{2})}{2\pi }\Omega _{\text{F}}$ (unimportant global shift
is ignored). Thanking to the topological protection, we find that our scheme
works very well as long as the energy shift is smaller than the smallest
band gap $\frac{|l|\delta \omega }{4J_{1}}\lesssim 1$, as confirmed by our
numerical simulation shown in Fig.~\ref{fig:error_a}. Without loss of
generality, we consider 
above rectangular-shaped cavity with $F=10$cm. $\frac{\delta \omega }{4J_{1}}
$ as a function of $X$ and $Y$ is shown in Fig.~\ref{fig:stable_a}, and $%
\frac{|l|\delta \omega }{4J_{1}}\lesssim 1$ leads to $|l|\sqrt{(X-F)(Y-F)}%
\lesssim 1000\mu $m. 
If we encode the quantum information in the space $l\in \lbrack -l_{\text{max%
}},l_{\text{max}}]$ (i.e., the maximal OAM to be switched to is $|l|=l_{%
\text{max}}$), our scheme works well when $l_{\text{max}}\lesssim \frac{%
1000\mu \text{m}}{\sqrt{(X-F)(Y-F)}}$. Using standard micrometric linear
translation stages, one can easily achieve an accuracy with $(X-F)\sim
(Y-F)\sim \mu $m, thus corresponding $l_{\text{max}}$ can be up to hundreds.
Nanometer-precision linear translation stage (e.g. M-714.2HD, PI, Germany)
can reach an accuracy of 0.05$\mu $m, which yields a maximal usable OAM of
several thousands in principle.


\begin{thebibliography}{99}
\bibitem{nielsen2002quantum} M.~A.~Nielsen and I.~L.~Chuang,
\newblock {\em
Quantum Computation and Quantum Information} (Cambridge University Press,
Cambridge, 2000).

\bibitem{gisin2002quantum} N.~Gisin, G.~Ribordy, W.~Tittel, and H.~Zbinden, %
\newblock Quantum cryptography, \newblock \href{http://dx.doi.org/10.1103/RevModPhys.74.145}%
{Rev. Mod. Phys. \textbf{74}, 145 (2002)}.

\bibitem{culcer2012valley} D.~Culcer, A.~L.~Saraiva, B.~Koiller, X.~Hu, and
S.~Das Sarma, \newblock Valley-based noise-resistant quantum computation
using si quantum dots, \newblock \href{http://dx.doi.org/10.1103/PhysRevLett.108.126804}%
{Phys. Rev. Lett. \textbf{108}, 126804 (2012)}.

\bibitem{andersen2015hybrid} U.~L.~Andersen, J.~S.~Neergaard-Nielsen,
P.~Van~Loock, and A.~Furusawa, \newblock Hybrid discrete-and
continuous-variable quantum information, \newblock \href{http://dx.doi.org/10.1038/nphys3410}%
{Nature Phys. \textbf{11}, 713 (2015)}.

\bibitem{allen1992orbital} L.~Allen, M.~W.~Beijersbergen, R.~J.~C.~Spreeuw,
and J.~P.~Woerdman, \newblock Orbital angular momentum of light and the
transformation of laguerre-gaussian laser modes, \newblock \href{http://dx.doi.org/10.1103/PhysRevA.45.8185}%
{Phys. Rev. A \textbf{45}, 8185 (1992)}.

\bibitem{molina2007twisted} G.~Molina-Terriza, J.~P.~Torres, and L.~Torner, %
\newblock Twisted photons, \newblock \href{http://dx.doi.org/10.1038/nphys607}%
{Nature Phys. \textbf{3}, 305 (2007)}.

\bibitem{yao2011orbital} A.~M.~Yao and M.~J.~Padgett, \newblock Orbital
angular momentum: origins, behavior and applications, \newblock \href{http://dx.doi.org/10.1364/AOP.3.000161}%
{Adv. Opt. Photon. \textbf{3}, 161 (2011)}.

\bibitem{malik2016multi} M.~Malik, M.~Erhard, M.~Huber, M.~Krenn,
R.~Fickler, and A.~Zeilinger, \newblock Multi-photon entanglement in high
dimensions, \newblock \href{http://dx.doi.org/10.1038/nphoton.2016.12}{%
Nature Photon. \textbf{10}, 248 (2016)}.

\bibitem{barreiro2008beating} J.~T.~Barreiro, T.-C.~Wei, and P.~G.~Kwiat, %
\newblock Beating the channel capacity limit for linear photonic superdense
coding, \newblock \href{http://dx.doi.org/10.1038/nphys919}{Nature Phys.
\textbf{4}, 282 (2008)}.

\bibitem{wang2012terabit} J.~Wang, \emph{et al.},
\newblock Terabit free-space data transmission employing orbital angular
momentum multiplexing, \newblock \href{http://dx.doi.org/10.1038/nphoton.2012.138}%
{Nature Photon. \textbf{6}, 488 (2012)}.

\bibitem{cardano2015quantum} F.~Cardano, \emph{et al.},
\newblock Quantum walks and wavepacket dynamics on a lattice with twisted
photons, \newblock \href{http://dx.doi.org/10.1126/sciadv.1500087}{Sci. Adv.
\textbf{1}, e1500087 (2015)}.

\bibitem{cardano2016statistical} F.~Cardano, \emph{et al.},
\newblock Statistical moments of quantum-walk dynamics reveal topological
quantum transitions, \newblock \href{http://dx.doi.org/10.1038/ncomms11439}{%
Nature Commun. \textbf{7}, 11439 (2016)}.

\bibitem{luo2015quantum} X.-W.~Luo, X.~Zhou, C.-F.~Li, J.-S.~Xu, G.-C.~Guo,
and Z.-W.~Zhou, \newblock Quantum simulation of 2d topological physics in a
1d array of optical cavities, \newblock \href{http://dx.doi.org/10.1038/ncomms8704}%
{Nature Commun. \textbf{6}, 7704 (2015)}.

\bibitem{zhou2017dynamically} X.-F.~Zhou, X.-W.~Luo, S.~Wang, G.-C.~Guo, X.~Zhou, H.~Pu, and Z.-W.~Zhou,
\newblock Dynamically Manipulating Topological Physics and Edge Modes in a Single Degenerate Optical Cavity,
\newblock \href{http://dx.doi.org/10.1103/PhysRevLett.118.083603}{%
Phys. Rev. Lett. \textbf{118}, 083603 (2017)}.

\bibitem{luo2017synthetic} X.-W.~Luo, X.~Zhou, C.-F.~Li, J.-S.~Xu, G.-C.~Guo, C.~Zhang,
and Z.-W.~Zhou, \newblock Synthetic-lattice enabled all-optical devices based on orbital angular momentum of light, \newblock \href{http://dx.doi.org/10.1038/ncomms16097}%
{Nature Commun. \textbf{8}, 16097 (2017)}.

\bibitem{fickler2012quantum} R.~Fickler, R.~Lapkiewicz, W.~N.~Plick,
M.~Krenn, C.~Schaeff, S.~Ramelow, and A.~Zeilinger, \newblock Quantum
entanglement of high angular momenta, \newblock \href{http://dx.doi.org/10.1126/science.1227193}%
{Science \textbf{338}, 640 (2012)}.

\bibitem{fickler2014interface} R.~Fickler, R.~Lapkiewicz, M.~Huber,
M.~P.~J.~Lavery, M.~J.~Padgett, and A.~Zeilinger, \newblock Interface
between path and orbital angular momentum entanglement for high-dimensional
photonic quantum information, \newblock \href{http://dx.doi.org/10.1038/ncomms5502}%
{Nature Commun. \textbf{5}, 2014}.

\bibitem{wang2015quantum} X.-L.~Wang, X.-D.~Cai, Z.-E.~Su, M.-C.~Chen,
D.~Wu, L.~Li, N.-L.~Liu, C.-Y.~Lu, and J.-W.~Pan, \newblock Quantum
teleportation of multiple degrees of freedom of a single photon, \newblock
\href{http://dx.doi.org/10.1038/nature14246}{Nature, \textbf{518}, 516 (2015)%
}.

\bibitem{groblacher2006experimental} S.~Gr{\"o}blacher, T.~Jennewein,
A.~Vaziri, G.~Weihs, and A.~Zeilinger, \newblock Experimental quantum
cryptography with qutrits, \newblock \href{http://dx.doi.org/10.1088/1367-2630/8/5/075}%
{New J. Phys. \textbf{8}, 75 (2006)}.

\bibitem{mafu2013higher} M.~Mafu, \emph{et al.},
\newblock Higher-dimensional orbital-angular-momentum-based quantum key
distribution with mutually unbiased bases, \newblock \href{http://dx.doi.org/10.1103/PhysRevA.88.032305}%
{Phys. Rev. A \textbf{88}, 032305 (2013)}.

\bibitem{vallone2014free} G.~Vallone, \emph{et al.},
\newblock Free-space quantum key distribution by rotation-invariant twisted
photons, \newblock \href{http://dx.doi.org/10.1103/PhysRevLett.113.060503}{%
Phys. Rev. Lett. \textbf{113}, 060503 (2014)}.

\bibitem{mirhosseini2015high} M.~Mirhosseini, \emph{et al.},
\newblock High-dimensional quantum cryptography with twisted light, %
\newblock \href{http://dx.doi.org/:10.1088/1367-2630/17/3/033033}{New J.
Phys. \textbf{17}, 033033 (2015)}.

\bibitem{thalhammer2013speeding} G.~Thalhammer, R.~W.~Bowman, G.~D.~Love,
M.~J.~Padgett, and M.~Ritsch-Marte, \newblock Speeding up liquid crystal
slms using overdrive with phase change reduction, \newblock \href{http://dx.doi.org/10.1364/OE.21.001779}%
{Opt. Exp. \textbf{21}, 1779 (2013)}.

\bibitem{mirhosseini2013rapid} M.~Mirhosseini, O.~S.~Magana-Loaiza, C.~Chen,
B.~Rodenburg, M.~Malik, and R.~W.~Boyd, \newblock Rapid generation of light
beams carrying orbital angular momentum, \newblock \href{http://dx.doi.org/10.1364/OE.21.030196}%
{Opt. Exp. \textbf{21}, 30196 (2013)}.

\bibitem{radwell2014high} N.~Radwell, D.~Brickus, T.~W.~Clark, and
S.~Franke-Arnold, \newblock High speed switching between arbitrary spatial
light profiles, \newblock \href{http://dx.doi.org/10.1364/OE.22.012845}{Opt.
Exp. \textbf{22}, 12845 (2014)}.

\bibitem{strain2014fast} M.~J.~Strain, \emph{et al.},
\newblock Fast electrical switching of orbital angular momentum modes using
ultra-compact integrated vortex emitters, \newblock \href{http://dx.doi.org/10.1038/ncomms5856}%
{Nature Commun. \textbf{5}, 4856 (2014)}.

\bibitem{marrucci2006optical} L.~Marrucci, C.~Manzo, and D.~Paparo, %
\newblock Optical spin-to-orbital angular momentum conversion in
inhomogeneous anisotropic media, \newblock \href{http://dx.doi.org/10.1103/PhysRevLett.96.163905}%
{Phys. Rev. Lett. \textbf{96}, 163905 (2006)}.

\bibitem{slussarenko2011efficient} S.~Slussarenko, E.~Karimi, B.~Piccirillo,
L.~Marrucci, and E.~Santamato. \newblock Efficient generation and control of
different-order orbital angular momentum states for communication links, %
\newblock \href{http://dx.doi.org/10.1364/JOSAA.28.000061}{J. Opt. Soc. Am.
A \textbf{28}, 61 (2011)}.

\bibitem{haldane2008possible} F. D. M.~Haldane, and S.~Raghu, \newblock %
Possible Realization of Directional Optical Waveguides in Photonic Crystals
with Broken Time-Reversal Symmetry, \newblock \href{http://dx.doi.org/10.1103/PhysRevLett.100.013904}%
{Phys. Rev. Lett. \textbf{100}, 013904 (2008)}.

\bibitem{hafezi2011robust} M.~Hafezi, E.~A.~Demler, M.~D.~Lukin, and
J.~M.~Taylor, \newblock Robust optical delay lines with topological
protection, \newblock \href{http://dx.doi.org/10.1038/nphys2063}{Nature
Phys. \textbf{7}, 907 (2011)}.

\bibitem{fang2012realizing} K.~Fang, Z.~Yu, and S.~Fan, \newblock Realizing
effective magnetic field for photons by controlling the phase of dynamic
modulation, \newblock \href{http://dx.doi.org/10.1038/nphoton.2012.236}{%
Nature Photon. \textbf{6}, 782 (2012)}.

\bibitem{lin2014light} Q.~Lin, and S.~Fan, \newblock Light guiding by
effective gauge field for photons, \newblock \href{http://dx.doi.org/10.1103/PhysRevX.4.031031}%
{Phys. Rev. X \textbf{4}, 031031 (2014)}.

\bibitem{lu2014topological} L.~Lu, J.~D.~Joannopoulos, and M.~Solja{\v{c}}i{%
\'c}, \newblock Topological photonics, \newblock \href{http://dx.doi.org/10.1038/nphoton.2014.248}%
{Nature Photon. \textbf{8}, 821 (2014)}.

\bibitem{yuan2016photonic} L.~Yuan, Y.~Shi, and S.~Fan, \newblock Photonic
gauge potential in a system with a synthetic frequency dimension, \newblock
\href{http://dx.doi.org/10.1364/OL.41.000741}{Opt. Lett. \textbf{41}, 741
(2016)}.

\bibitem{lin2016photonic} Q.~Lin, M.~Xiao, L.~Yuan, and S.~Fan, \newblock %
Photonic Weyl point in a two-dimensional resonator lattice with a synthetic
frequency dimension, \newblock \href{http://dx.doi.org/10.1038/ncomms13731}{%
Nature Commun. \textbf{7}, 13731 (2016)}.

\bibitem{thouless1983quantization} D.~J.~Thouless, \newblock Quantization of
particle transport, \newblock \href{http://dx.doi.org/10.1103/PhysRevB.27.6083}%
{Phys. Rev. B \textbf{27}, 6083 (1983)}.

\bibitem{lohse2015thouless} M.~Lohse, C.~Schweizer, O.~Zilberberg,
M.~Aidelsburger, and I.~Bloch, \newblock A thouless quantum pump with
ultracold bosonic atoms in an optical superlattice, \newblock \href{http://dx.doi.org/10.1038/nphys3584}%
{Nature Phys. \textbf{12}, 350 (2016)}.

\bibitem{nakajima2016topological} S.~Nakajima, \emph{et al.},
\newblock Topological thouless pumping of ultracold fermions, \newblock
\href{http://dx.doi.org/10.1038/nphys3622}{Nature Phys. \textbf{12}, 296
(2016)}.

\bibitem{kraus2012topological} Y.~E.~Kraus, Y.~Lahini, Z.~Ringel, M.~Verbin,
and O.~Zilberberg, \newblock Topological states and adiabatic pumping in
quasicrystals, \newblock \href{http://dx.doi.org/10.1103/PhysRevLett.109.106402}%
{Phys. Rev. Lett. \textbf{109}, 106402 (2012)}.

\bibitem{verbin2015topological} M.~Verbin, O.~Zilberberg, Y.~Lahini,
Y.~E.~Kraus, and Y.~Silberberg, \newblock Topological pumping over a
photonic fibonacci quasicrystal, \newblock \href{http://dx.doi.org/10.1103/PhysRevB.91.064201}%
{Phys. Rev. B \textbf{91}, 064201 (2015)}.

\bibitem{celi2014synthetic} A.~Celi, \emph{et al.},
\newblock Synthetic gauge fields in synthetic dimensions, \newblock \href{http://dx.doi.org/10.1103/PhysRevLett.112.043001}%
{Phys. Rev. Lett. \textbf{112}, 043001 (2014)}.

\bibitem{mancini2015observation} M.~Mancini, \emph{et al.}, \newblock %
Observation of chiral edge states with neutral fermions in synthetic hall
ribbons, \newblock \href{http://dx.doi.org/10.1126/science.aaa8736}{Science
\textbf{349}, 1510 (2015)}.

\bibitem{stuhl2015visualizing} B.~K.~Stuhl, H.-I.~Lu, L.~M.~Aycock,
D.~Genkina, and I.~B.~Spielman, \newblock Visualizing edge states with an
atomic bose gas in the quantum hall regime, \newblock \href{http://dx.doi.org/10.1126/science.aaa8515}%
{Science \textbf{349}, 1514 (2015)}.

\bibitem{arnaud1969degenerate} J.~A.~Arnaud, \newblock Degenerate optical
cavities, \newblock \href{http://dx.doi.org/10.1364/AO.8.000189}{Appl. Opt.
\textbf{8}, 189 (1969)}.

\bibitem{horak2002optical} P.~Horak, \emph{et al.}, \newblock Optical
kaleidoscope using a single atom, \newblock \href{http://dx.doi.org/10.1103/PhysRevLett.88.043601}%
{Phys. Rev. Lett. \textbf{88}, 043601 (2002)}.

\bibitem{gopalakrishnan2009emergent} S.~Gopalakrishnan, B.~L.~Lev, and
P.~M.~Goldbart, \newblock Emergent crystallinity and frustration with
Bose--Einstein condensates in multimode cavities, \newblock \href{http://dx.doi.org/10.1038/nphys1403}%
{Nature Phys. \textbf{5}, 845 (2009)}.

\bibitem{ballantine2017meissner} K.~E.~Ballantine, B.~L.~Lev, and
J.~Keeling, \newblock Meissner-like Effect for a Synthetic Gauge Field in
Multimode Cavity QED, \newblock \href{http://dx.doi.org/10.1103/PhysRevLett.118.045302}%
{Phys. Rev. Lett. \textbf{118}, 045302 (2017)}.

\bibitem{schine2016synthetic} N.~Schine, A.~Ryou, A.~Gromov, A.~Sommer, and
J.~Simon, \newblock Synthetic Landau levels for photons, \newblock \href{http://dx.doi.org/10.1038/nature17943}%
{Nature, \textbf{534}, 671 (2016)}.

\bibitem{yariv2007photonics} A.~Yariv and P.~Yeh,
\newblock {\em Photonics:
Optical Electronics in Modern Communications} (Oxford University Press,
Oxford, 2007).

\bibitem{hodgson2005laser} N.~Hodgson and H.~Weber,
\newblock {\em Laser
Resonators and Beam Propagation: Fundamentals, Advanced Concepts,
Applications} (Springer, New York, 2005).


\bibitem{oemrawsingh2005experimental} S.~S.~R.~Oemrawsingh, \emph{et al.},
\newblock Experimental demonstration of fractional orbital angular momentum
entanglement of two photons, \newblock \href{http://dx.doi.org/10.1103/PhysRevLett.95.240501}%
{Phys. Rev. Lett. \textbf{95}, 240501 (2005)}.

\bibitem{karimi2009efficient} E.~Karimi, B.~Piccirillo, E.~Nagali,
L.~Marrucci, and E.~Santamato, \newblock Efficient generation and sorting of
orbital angular momentum eigenmodes of light by thermally tuned q-plates, %
\newblock \href{http://dx.doi.org/10.1063/1.3154549}{Appl. Phys. Lett.
\textbf{94}, 231124 (2009)}.

\bibitem{leach2002measuring} J.~Leach, M.~J.~Padgett, S.~M.~Barnett,
S.~Franke-Arnold, and J.~Courtial, \newblock Measuring the orbital angular
momentum of a single photon, \newblock \href{http://dx.doi.org/10.1103/PhysRevLett.88.257901}%
{Phys. Rev. Lett. \textbf{88}, 257901 (2002)}.

\bibitem{ganeshan2015constructing} S.~Ganeshan and S.~Das Sarma. \newblock %
Constructing a weyl semimetal by stacking one-dimensional topological
phases, \newblock \href{http://dx.doi.org/10.1103/PhysRevB.91.125438}{Phys.
Rev. B \textbf{91}, 125438 (2015)}.

\bibitem{rice1982elementary} M.~J.~Rice and E.~J.~Mele, \newblock Elementary
excitations of a linearly conjugated diatomic polymer, \newblock \href{http://dx.doi.org/10.1103/PhysRevLett.49.1455}%
{Phys. Rev. Lett. \textbf{49}, 1455 (1982)}.

\bibitem{walls2008quantum} D.~F.~Walls and G.~J.~Milburn,
\newblock {\em
Quantum optics} (Springer-Verlag, Berlin, 2008).

\bibitem{hernandez1986fabry} G.~Hern{\'a}ndez,
\newblock {\em Fabry-perot
interferometers} \newblock (Cambridge University Press, Cambridge, 1986).

\bibitem{PhysRevLett.101.190501} X.~Bao, Y.~Qian, J.~Yang, H.~Zhang, Z.~Chen, T.~Yang, J.-W.~Pan,
\newblock Generation of Narrow-Band Polarization-Entangled Photon Pairs for Atomic Quantum Memories, \newblock \href{http://dx.doi.org/10.1103/PhysRevLett.101.190501}%
{Phys. Rev. Lett. \textbf{101}, 190501 (2008)}.

\bibitem{nagorny2003collective} B.~Nagorny, Th.~Els{\"a}sser, and
A.~Hemmerich, \newblock Collective atomic motion in an optical lattice
formed inside a high finesse cavity, \newblock \href{http://dx.doi.org/10.1103/PhysRevLett.91.153003}%
{Phys. Rev. Lett. \textbf{91}, 153003 (2003)}.

\bibitem{marrucci2011spin} L.~Marrucci, \emph{et al.},
\newblock Spin-to-orbital conversion of the angular momentum of light and
its classical and quantum applications, \newblock \href{http://dx.doi.org/10.1088/2040-8978/13/6/064001}%
{J. Opt. \textbf{13}, 064001 (2011)}.

\bibitem{raut2011anti} H.~K.~Raut, V.~A.~Ganesh, A.~S.~Nair, and
S.~Ramakrishna,
\newblock Anti-reflective coatings: A critical, in-depth
review, \newblock \href{http://dx.doi.org/10.1039/C1EE01297E}{Energy Environ. Sci. \textbf{4}, 3779--3804 (2011)}.
\end{thebibliography}

\end{document}